\documentclass{article}
\usepackage{amsfonts,amssymb, amsmath}
\usepackage[english]{babel}

\DeclareMathAlphabet{\mathpzc}{OT1}{pzc}{m}{it}

\def\nn{\nonumber }
\def\bq{ \begin{equation} }
\def\eq{ \end{equation} }
\def\ben{ \begin{eqnarray} }
\def\en{ \end{eqnarray} }

\def\ii{{\rm i}}
\def\ee{{\rm e}}

\newtheorem{prop}{Proposition}

\newtheorem{re}{Remark}
\newenvironment{rem}{\begin{re} \rm }{\end{re}}
\begin{document}


\title{On the generalized integrable Chaplygin system. }
\author{ A V Tsiganov \\
\it\small
St.Petersburg State University, St.Petersburg, Russia\\
\it\small e--mail: andrey.tsiganov@gmail.com}

\date{}
\maketitle

\begin{abstract}
We discuss two polynomial bi-Hamiltonian structures for the generalized integrable Chaplygin system on the sphere ${\mathcal S}^2$ with an additional integral of fourth order in momenta. An explicit procedure to find the variables of separation, the separation relations and the transformation of the corresponding algebraic curves of genus two is considered in detail.
\end{abstract}


\vskip0.1truecm

\section{Introduction}
\setcounter{equation}{0}
We address the problem of the separation of variables for the Hamilton-Jacobi equation within the theoretical scheme of bi-hamiltonian geometry. The main aim is the construction of  different variables of separation for the given integrable system without any additional information (Lax matrices, $r$-matrices, links with soliton equations  etc.)

Different variables of separation may be useful in different perturbation theories,as well as distinct procedures of quantization and various methods of qualitative analysis etc. From mathematical point of view the notion of different variables of separation allows us to study relations between distinct algebraic curves associated with the corresponding separated equations. Such relations give us a lot of new examples of reductions of Abelian integrals and, therefore, they may be a source of new ideas in number theory, algebraic geometry and modern cryptography.

The paper is organized as follows. In Section II, the necessary aspects of bi-hamiltonian geometry are briefly reviewed. Then, we discuss a possible application of these methods to calculation of the polynomial bi-hamiltonian structures for the given Chaplygin system. In Section III, the problem of finding  variables of separation and  corresponding separation relations is treated and solved. Some applications of  real and complex variables of separation are discussed in the final Section.

\section{The bi-hamiltonian structure of the Chaplygin system}
\setcounter{equation}{0}
In order to get variables of separation according to  general usage of bi-hamiltonian geometry  firstly  we have to calculate the bi-hamiltonian structure for the given integrable system with integrals of motion $H_1,\ldots,H_n$ on the Poisson manifold $M$ with the kinematic Poisson bivector $P$ and the Casimir functions $C_1,\ldots,C_k$ \cite{fp02,mag97,ts08,ts08b,ts09}.

Let us consider the generalized Chaplygin system \cite{chap,kuzts89,ts02} with the following integrals of motion
\ben\label{H1-0}
H_1&=& J_1^2+J_2^2+2J_3^2+c_2(x_1^2-x_2^2) +\dfrac{c_4}{x_3^2},\qquad c_2,c_4\in\mathbb R,\\
H_2&=&\left(J_1^2+J_2^2+\dfrac{c_4}{x_3^2}\right)^2+2c_2x_3^2(J_1^2-J_2^2)+c_2^2x_3^4.\nn
\en
In this standard coordinates $J=(J_1,J_2,J_3)$ and $ x=(x_1,x_2,x_3)$ on
the Euclidean algebra $e^*(3)$ the Poisson bivector looks like the following antisymmetric matrix
\[
P=\left(
 \begin{array}{cccccc}
 0 & 0 & 0 & 0 & x_3 & -x_2 \\
 * & 0 & 0 & -x_3 & 0 & x_1 \\
 * & * & 0 & x_2 & -x_1 & 0 \\
 * & * & * & 0 & J_3 & -J_2 \\
 * & * & * & * & 0 & J_1 \\
 * & * & * & * & * & 0 \\
 \end{array}
\right).
\]
It has two Casimir elements
\bq \label{caz-e3}
PdC_{1,2}=0,\qquad C_1=a^2\equiv\sum_{k=1}^3 x_k^2, \qquad C_2= \langle x,J \rangle\equiv\sum_{k=1}^3 x_kJ_k .\eq
At $C_2=\langle x,J\rangle=0$ this integrals of motion are in  involution
\[
\{H_1,H_2\}=\langle PdH_1,dH_2\rangle=0.
\]
and the corresponding symplectic leaves are equivalent to cotangent bundle $T^*{\mathcal S}^2$ of the sphere with the radius $a=|x|$. We will consider such symplectic leaves only, and, therefore, all the formulae below hold true up to $C_2=0$.

Following to \cite{ts07c,ts08,ts08b,ts09} we will suppose that the desired second Poisson bivector is
the Lie derivative of $P$ along some unknown Liouville vector field $X$
\bq\label{co-b}
P'=\mathcal L_X(P)
\eq
which has to satisfy to the equations
\bq\label{m-eq12}
\{H_1,H_2\}'=\langle P'dH_1,dH_2\rangle=0,\qquad [P',P']=[\mathcal L_X(P),\mathcal L_X(P)]=0,
\eq
where $[.,.]$ means the Schouten bracket.

Obviously enough, in their full generality  equations (\ref{m-eq12}) are too difficult to be solved because it has infinitely many solutions \cite{ts07a,ts08b}. In order to get some particular solutions we will use the  additional assumption
\bq\label{m-eq3}
P'dC_{1,2}=0,
\eq
and polynomial in momenta $J$ \textit{ans\"{a}tze} for the components $X^j$, $j=1,\ldots,6$,of the Liouville vector field $X=\sum X^j\,\partial_j$ \cite{ts09,ts07c,ts08}.

Substituting this \textit{ans\"{a}tze} into the equations (\ref{m-eq12}-\ref{m-eq3}) and demanding that all the coefficients at powers of $J$ vanish one gets the over determined system of algebro-differential equations on functions of $x$ which can be easily solved in the modern computer algebra systems.

For the linear ans\"{a}tze one gets the trivial solutions $P'=P$ only. The quadratic ans\"{a}tze yields first non-trivial solution
\[
X=\dfrac{W_3}{a^2x_3}\Bigl(W_1,W_2,W_3,0,0,0\Bigr)-\dfrac{c_2J_1}{x_3}\Bigl(0,0,0,x_3,0,-x_1\Bigr),\qquad W=J\times x,
\]
where $W$ is a cross product of $J$ and $x$. This Liuville vector field gives rise to the following real Poisson bivector
\bq\label{P2}
P'=\left(
 \begin{array}{cccccc}
 0 & J_3 & -\frac{x_2J_3}{x_3} & \frac{J_1W_1}{x_3^2} & \frac{J_2W_1}{x_3^2} & 0 \\
 * & 0 & \frac{x_1J_3}{x_3} & \frac{J_1W_2}{x_3^2} & \frac{J_2W_2}{x_3^2} & 0 \\
 * & * & 0 & \frac{J_1W_3}{x_3^2} & \frac{J_2W_3}{x3^2} & 0 \\
 * & * & * & 0 & 0 & 0 \\
 * & * & * & * & 0 & 0 \\
 * & * & * & * & * & 0
 \end{array}
\right)-c_2
\left(
 \begin{array}{cccccc}
 0 & 0 & 0 & 0 & 0 & 0 \\
 * & 0 & 0 & x_3 & 0 & -x_1 \\
 * & * & 0 &-x_2 & 0 & \frac{x_1 x_2}{x_3} \\
 * & * & * & 0 & \frac{x_2J_2}{x_3} & \frac{x_3^2J_2+x_1x_2J_1}{x_3^2} \\
 * & * & * & * & 0 & \frac{x_1x_2J_2}{x_3^2} \\
 * & * & * & * & * & 0
 \end{array}
\right),
\eq

For the cubic ans\"{a}tze we get one real Poisson bivector compatible with (\ref{P2})
and two new hermitian conjugated bivectors $P'$ and $P'^*$. This new and more cumbersome solution
reads as
\[
P'_{12}=x_3(W_3+\ii x_3J_3),\qquad P'_{1,3}=-x_2(W_3+\ii x_3J_3),\qquad P'_{2,3}=x_1(W_3+\ii x_3J_3)\]
\ben
P'_{14}&=&J_2W_1-\ii x_3J_1J_3-\sqrt{c_2}x_3(x_2J_3-W_1)+\dfrac{x_2^2(c_2x_3^4-c_4)}{x_3^3}\nn\\
P'_{15}&=&-(J_1+\ii J_2)W_1-\frac{\ii x_3}{2}(J_1^2-J_2^2)
+\ii\sqrt{c_2}(x_2J_3-W_1)x_3-\frac{(\ii x_3^2-2x_1x_2)(c_2x_3^4+c_4)}{2x_3^3}\nn\\
P'_{16}&=&-\ii J_3W_1+\frac{\ii x_2(T-\ii c_2
(\ii x_3^2-4x_1x_2))}{2}-\sqrt{c_2}(W_1-x_2J_3)(x_1-\ii x_2)+\frac{\ii c_4x_2}{2x_3^2}\nn\\
P'_{24}&=&-\ii(J_1+\ii J_2)W_2-\frac{x_3}{2}(J_1^2-J_2^2)+\sqrt{c_2}x_3(x_1J_3+W_2)-\frac{(\ii x_3^2+2x_1x_2)(c_2x_3^4-c_4)}{2x_3^3}\nn\\
P'_{25}&=&-J_1W_2-\ii x_1J_2J_3-\ii\sqrt{c_2}x_3(x_1J_3+W_2)-\frac{x_1^2(c_2x_3^4+c_4)}{x_3^3}\label{P3}\\
P'_{26}&=&-\ii J_3W_2-\dfrac{\ii x_1 (T-\ii c_2(\ii x_3^2+4x_1x_2)}{2}-\sqrt{c_2}(x_1J_3+W_2)(x_1-\ii x_2)-\dfrac{\ii c_4x_1}{x_3^2}\nn\\
P'_{34}&=&J_2W_3+\frac{\ii x_2(J_1^2+J_2^2)}{2}+\sqrt{c_2}x_3W_3+\frac{\ii x_2(c_2x_3^4-c_4)}{2x_3^2}\nn\\
P'_{35}&=&-J_1W_3-\frac{\ii x_1(J_1^2+J_2^2)}{2}-\ii\sqrt{c_2}x_3W_3+\dfrac{\ii x_1(c_2x_3^4+c_4)}{2x_3^2}\nn\\
P'_{36}&=&-\ii J_3W_3-\sqrt{c_2}W_3(x_1-\ii x_2)-\ii c_2x_1x_2x_3\nn
\en
\ben
P'_{45}&=&\frac{\ii(J_1^2+J_2^2)J_3}{2}+\sqrt{c_2}\left(
c_2x_3^2(x_1+\ii x_2)+\frac{(x_1-\ii x_2)c_4}{x_3^2}-x_3J_3(J_1-\ii J_2)\right)\nn\\
&-&\frac{\ii x_1(J_1+2\ii J_2)(c_2x_3^4+c_4)}{2x_3^3}
-\frac{\ii x_2(J_2+2\ii J_1)(c_2x_3^4-c_4)}{2x_3^3}\nn\\
P'_{46}&=&\frac{\ii J_2 T}{2}-\sqrt{c_2}\left(c_2x_2x_3(x_1+\ii x_2)-(x_1-\ii x_2)\Bigl(J_2J_3-\frac{c_4x_2}{x_3^3}\Bigr)\right)\nn\\
&-&x_1x_2\left(2 c_2J_2+\frac{\ii c_4J_1}{x_3^4}\right)-\frac{\ii J_2}{2}(x_3^2+2x_2^2)\left(c_2-\dfrac{c_4}{x_3^4}\right)\nn\\
P'_{56}&=&-\dfrac{\ii J_1T}{2}+\sqrt{c_2}\left(
c_1x_1x_3(x_1+\ii x_2)-(x_1-\ii x_2)\Bigl(J_1J_3-\dfrac{c_4x_1}{x_3^3}\Bigr)\right)\nn\\
&+&x_1x_2\left(2c_2J_1-\frac{\ii c_4J_2}{x_3^4}\right)-\dfrac{\ii J_1}{2}(x_3^2+2x_1^2)\left(c_2+\frac{c_4}{x_3^4}\right)\,.\nn
\en
Here $T=J_1^2+J_2^2+2J_3^2$ and $\ii=\sqrt{-1}$. This cubic Poisson structure may be rewritten in lucid form by using $2\times 2$ Lax matrices for the Chaplygin system \cite{kuzts89,ts02} and the bi-hamiltonian structure associated with the reflection equation algebra \cite{ts08c}.

The quartic ans\"{a}tze yields a lot of solutions, which will be classified and
studied at a future date.

To sum up, using applicable ans\"{a}tze for the Liouville vector field $X$ we get a real relatively simple quadratic bivector (\ref{P2}) and a more complicated complex cubic bivector (\ref{P3}). Modern computer software allows to do it on a personal computer wasting only few seconds. The application of this Poisson bivectors will be given in the next section.

\section{Variables of separation and separation relations}
\setcounter{equation}{0}
The second step in the bi-hamiltonian method of separation of variables is calculation of canonical variables of separation $(q_1,\dots,q_n,p_1,\dots,p_n)$ and separation relations of the form
\begin{equation}
\label{seprelint}
\phi_i(q_i,p_i,H_1,\dots,H_n)=0\ ,\quad i=1,\dots,n\ ,
\qquad\mbox{with }\det\left[\frac{\partial \phi_i}{\partial H_j}\right]
\not=0\> .
\end{equation}
The reason for this definition is that the stationary
Hamilton-Jacobi equations for the Hamiltonians $H_i$ can be
collectively solved by the additively separated complete integral
\begin{equation}\label{eq:i0}
W(q_1,\dots,q_n;\alpha_1,\dots,\alpha_n)=
\sum_{i=1}^n W_i(q_i;\alpha_1,\dots,\alpha_n)\>,
\end{equation}
where  $W_i$ are found by quadratures as  solutions of ordinary
differential equations.

According to \cite{fp02,ts08,ts09},  separated coordinates $q_j$ are the eigenvalues of the control matrix $F$ defined by \[
P'{\mathbf{dH}}=P\bigl(F{\mathbf{dH}}\bigr).
\]
Its eigenvalues coincide with the Darboux-Nijenhuis coordinates (eigenvalues of the recursion operator) on the corresponding symplectic leaves. Using control matrix $F$ we can avoid the procedure of restriction of the bivectors $P$ and $P'$ on symplectic leaves, that is a necessary intermediate calculation for the construction of the recursion operator \cite{fp02}.

Recall that independent integrals of motion $(H_1,\dots,H_n)$ are St\"ackel separable if the corresponding separation relations are given by the affine equations in  $H_j$, that is,
\begin{equation}
\label{stseprel}
\sum_{j=1}^n S_{ij}(p_i,q_i) H_j-U_i(p_i,q_i)=0\ ,\qquad
i=1,\dots,n\ ,
\end{equation}
with $S$  being an invertible matrix. Functions $S_{ij}$ and
$U_i$ depend only on one pair $(p_i,q_i)$ of  canonical variables of separation, thus 
it means that
\bq\label{st-mat}
\{S_{ik}, q_j\}=\{S_{ik}, p_j\}=\{ S_{ik},S_{jm} \}=0,\qquad i\neq j,
\eq
and 
\bq\label{st-pot}
\{U_{i}, q_j\}=\{U_{i}, p_j\}=\{ U_{i},U_{j} \}=0,\qquad i\neq j.
\eq
In this case $S$ is called a {\em St\"ackel matrix\/}, and $U$ - a {\em St\"ackel potential\/}.

For  St\"ackel separable systems the suitable normalized left eigenvectors of  control matrix $F$ form the St\"ackel matrix $S$ \cite{fp02}
 \[
 F=S^{-1}\,\mbox{diag}\,(q_1,\ldots,q_n)\,S.
 \]
However, we have to notice that the definition of  St\"ackel separability depends on the choice of $H_i$. Indeed, if
$(H_1,\dots,H_n)$ are St\"ackel-separable, then
$\widehat{H}_i=\widehat{H}_i(H_1,\dots,H_n)$ will not, in general,
fulfill relations of the form (\ref{stseprel}).

So, the main  problems are the finding of the conjugated momenta $p_{i}$, such that $\{p_i,q_j\}=\delta_{ij}$ and the construction of  separation relations $\phi_j$ (\ref{seprelint}) for generic non-St\"ackel separable systems.  We show below how we can solve these problems using the same control matrix $F$ with the addition of  some useful observations.

\subsection{The quadratic Poisson bivector}

According to \cite{fp02}, the bi-involutivity of integrals of motion
\[\{H_1,H_2\}=\{H_1,H_2\}'=0\]
is equivalent to the existence of the non-degenerate control matrix $F$ such that is
\bq\label{f-mat}
P'{\mathbf{dH}}=P\bigl(F{\mathbf{dH}}\bigr),\qquad\mbox{or}\qquad
P'dH_i=P\sum_{j=1}^2 F_{ij}\,dH_j,\qquad i=1,2.
\eq
For the quadratic in momenta Poisson bivector $P'$ (\ref{P2})  control matrix $F$ reads as
\bq
F=\left(
 \begin{array}{cc}
 \dfrac{1}{2}\left(\dfrac{J_1^2+J_2^2}{x_3^2}+c_2-\dfrac{c_4}{x_3^4}\right) & \dfrac{1}{4x_3^2} \\ \\
 \dfrac{(J_1^2+J_2^2)^2}{x_3^2}+2c_2(J_1^2-J_2^2)+c_2^2x_3^2-\dfrac{c_4^2}{x_3^6} & \dfrac12\left(\dfrac{J_1^2+J_2^2}{x_3^2}+c_2+\dfrac{c_4}{x_3^4}\right) \\
 \end{array}
\right)
\eq
The eigenvalues of this matrix are the required variables of separation $q_{1,2}$
\bq\label{q-def}
\det(F-\lambda I)=(\lambda-q_1)(\lambda-q_2)=\lambda^2-\left(\dfrac{J_1^2+J_2^2}{x_3^2}
+c_2\right)\lambda+\dfrac{c_2J_2^2}{x_3^2}.
\eq
If the corresponding separated relations are affine equations in $H_{1,2}$ then the suitable normalized left eigenvectors of $F$ form the St\"ackel matrix $S$
 \[
 F=S^{-1}\,\mbox{diag}\,(q_1,q_2)S.
 \]
In our case the matrix of normalized eigenvectors
\bq\label{gen-stm}
S=\left(
 \begin{array}{cc}
  s_1 & s_2 \\
  1 & 1 \\
 \end{array}
 \right),\qquad s_{1,2}=-\dfrac{2c_4}{x_3^2}\pm 2\sqrt{(J_1^2+J_2^2)^2+2c_2x_3^2(J_1^2-J_2^2)+c_2^2x_3^4\,}
\eq
does not form the standard St\"ackel matrix because
\[\{q_i,s_j\}\neq 0,\qquad\mbox{and}\qquad \{s_1,s_2\}\neq 0.\]
It means that the underlying separation relations do not form the St\"ackel affine equations (\ref{stseprel}) in $H_{1,2}$. Substituting functions on the integrals of motion $\widehat{H}_{1,2}=f_{1,2}(H_1,H_2)$
into the equation (\ref{f-mat})
\[P'{\mathbf{ d\widehat{H}}}=P\bigl(\widehat{F}{\mathbf{ d\widehat{H}}}\bigr)\,,\]
we can try to get a new control matrix $\widehat{F}$ which satisfies the St\"ackel properties (\ref{st-mat}). In the Chaplygin case it is a very simple calculation which yields to the following results
\[
\widehat{H}_1=H_1,\qquad \widehat{H}_2=H_2-H_1^2,
\]
and
\ben
\widehat{F}_{11}&=&    \frac{J_1^2+J_2^2+J_3^2}{x_3^2}+\frac{c_2(x_1^2-x_2^2+x_3^2)}{x_3^2},\qquad
\widehat{F}_{12}= \frac{1}{4x_3^2}, \nn\\
\widehat{F}_{21}&=&
     -\frac{4J_3^2(J_1^2+J_2^2+J_3^2)}{x_3^2}
     +2c_2\left(J_1^2-J_2^2-\frac{(x_1^2-x_2^2)(J_1^2+J_2^2+2J_3^2)}{x_3^2}\right)\nn\\
     &-&\frac{c_2^2(x_1^2-x_2^2+x_3^2)(x_1^2-x_2^2-x_3^2)}{x_3^2},\nn\\
\widehat{F}_{22}&=&-\frac{J_3^2}{x_3^2} -\frac{c_2(x_1^2-x_2^2-x_3^2)}{x_3^2},\nn
\en
so that
\[
\widehat{S}=\left(
 \begin{array}{cc}
  \widehat{s}_1 & \widehat{s}_2 \\
  1 & 1 \\

 \end{array}
 \right),\qquad \widehat{s}_{1,2}=2\left(\widehat{H}_1-\dfrac{c_4}{x_3^2}\right)\pm
 2\sqrt{(J_1^2+J_2^2)^2+2c_2x_3^2(J_1^2-J_2^2)+c_2^2x_3^4\,}\,.
\]
The St\"ackel conditions (\ref{st-mat}) are fulfilled and, therefore, $\widehat{s}_1$ is a function on the separated coordinate $q_1$ defined by (\ref{q-def}) and the unknown conjugated momenta $p_1$. It is easy to see that the recurrence chain
\bq\label{rrel1}
\phi_1=\{\widehat{s}_1,q_1\},\quad \phi_2=\{\phi_1,q_1\},\quad
\ldots,\quad \phi_i=\{\phi_{i-1},q_1\}
\eq
breaks down on the third step $\phi_3=0$. It means that $\widehat{s}_1$ is the second order polynomial in the momenta $p_1$ and, therefore, we can define this unknown momenta in the following way
\bq\label{p-def1}
p_1=\dfrac{\phi_1}{\phi_2}=-\dfrac{x_3}{2}\,\dfrac{(x_2J_1-x_1J_2)q_1+c_2x_1J_2}{(J_1^2+J_2^2)q_1-c_2J_2^2}
\eq
up to the canonical transformations $p_1\to p_1+g(q_1)$.

The similar calculation for $\widehat{s}_2$ yields definition of the second momenta
\bq\label{p-def2}
p_2=-\dfrac{x_3}{2}\,\dfrac{(x_2J_1-x_1J_2)q_2+c_2x_1J_2}{(J_1^2+J_2^2)q_2-c_2J_2^2}\,.
\eq
So, one gets canonical transformation from initial physical variables $x,J$ to the variables of separation $p,q$ (\ref{q-def},\ref{p-def1},\ref{p-def2}).

The inverse mapping looks like
\bq\label{inv-map}
J_1=\sqrt{q_1+q_2-c_2-\dfrac{q_1q_2}{c_2}}\,x_3,\quad J_2 =\sqrt{\dfrac{q_1q_2}{c_2}}\,x_3,\quad
J_3 =\dfrac{ 2\sqrt{q_1q_2(q_1-c_2)(c_2-q_2)}(p_1-p_2)}{q_1-q_2}
\eq
where $x_3=\sqrt{a^2-x_1^2-x_2^2}$ and
\[
x_1=2\sqrt{\dfrac{q_1q_2}{c_2}}\,
\dfrac{p_1q_1-p_2q_2-c_2(p_1-p_2)}{q_1-q_2},\qquad
x_2=-2\sqrt{\dfrac{(q_1-c_2)(c_2-q_2)}{c_2}}\,\dfrac{p_1q_1-p_2q_2}{q_1-q_2}\,.
\]
Using the elements of the St\"ackel matrix
\[\widehat{s}_{1,2}=2\Bigl(8(c_2-q_{1,2})\,q_{1,2}\,p_{1,2}^2-a^2(c_2-2q_{1,2})\Bigr)\,\]
 we can easily derive the required affine St\"ackel separated relations
\[\widehat{s}_k\widehat{H}_1+\widehat{H}_2=\dfrac{{\widehat s_k}^{\,2}}{4}-2c_4(c_2-2q_{k})\,,\qquad k=1,2.\]

If we come back to initial integrals of motion $H_{1,2}$ then
these separation relations go over to the equation
\ben\label{curve-2}
\mathcal C:\qquad \Phi(q,p)&=&\Bigl(8q (c_2-q )p ^2-a^2(c_2-2q )-H_1+\sqrt{H_2}\,\Bigr)
\\&\times&\Bigl(8q (c_2-q )p ^2-a^2(c_2-2q )-H_1-\sqrt{H_2}\,\Bigr)
-2c_4(c_2-2q )=0\,.\nn
\en
\begin{prop}
The variables of separation $(q_i,p_i)$ lie on the hyperelliptic algebraic curve $\mathcal C$ of the genus two $\mathrm g=2$ defined by (\ref{curve-2}) and the equations of motion are  linearized on its Jacobian.
\end{prop}

\begin{rem}
The separation relations (\ref{curve-2}) may be obtained in a framework of the St\"ackel formalizm (\ref{stseprel}) by using \textit{generalized} St\"ackel matrix $S$ (\ref{gen-stm}), whose
entries $S_{ij}$ depend on $(q_i,p_i)$ and on the integrals of motion
\[
s_i=\widehat{s}_{i}(p_i,q_i)-2H_1
\]
\end{rem}

\begin{rem}
At $c_4=0$ we reproduced the Chaplygin result so that  $(q_i,p_i)$ lie on a pair of  the elliptic curves of genus one
\bq\label{ell-curve}
\mathcal C_{1,2}:\qquad 8q_(c_2-q)p^2-a^2(c_2-2q)-H_1\pm\sqrt{H_2}=0\,.
\eq
Remind that every algebraic curve of genus one is isomorphic to a real torus.
\end{rem}

\subsection{The cubic Poisson bivector}
For the cubic in momenta Poisson bivector $P'$ (\ref{P3}) the entries of control matrix $F$ look like 
\ben
F_{11}&=&-\ii\,T+\sqrt{c_2}\bigl(x_3(J_1-\ii J_2)-2(x_1-\ii x_2)J_3\bigr)+2c_2x_1x_2-\dfrac{\ii c_4}{x_3^2},\nn\\
F_{21}&=&-\ii\left(J_1^2+J_2^2+\dfrac{c_4}{x_3^2}\right)\left(J_1^2+J_2^2+\dfrac{c_2}{x_3^2}+2\sqrt{c_2}x_3\ii(J_1-\ii J_2)\right)+\ii c_2^2x_3^4
\nn\\
&+&2c_2x_3^2\Bigl(\sqrt{c_2}x_3(J_1+\ii J_2)+2J_1J_2\Bigr)\,,\nn\\
F_{12}&=&\dfrac{\ii}{4},\qquad F_{22}=0\,.\nn
\en
The separated variables $\lambda_{1,2}$ are the roots of the characteristic polynomial
\bq\label{q-def2}
\det(F-\lambda I)=(\lambda-\lambda_1)(\lambda-\lambda_2)=\lambda^2-F_{1,1}\lambda-\dfrac{\ii}{4}\,F_{21}\,.
\eq
In the next step we have to find the conjugated momenta to these Darboux-Nijenhuis coordinates.

In \cite{kuzts89,ts02} we've found the separated coordinates $Q_{1,2}$ and the corresponding momenta $P_{1,2}$ for the Ko\-wa\-lev\-ski-Goryachev-Chaplygin gyrostat using $2\times 2$ Lax matrix, its Baker-Akhiezer vector-function and the reflection equation algebra. In the framework of Sklyanin formalism the separated coordinates are the poles of Baker-Akhiezer function with standard simplest normalization, whereas the separated momenta are expressed through the values of the elements of the Lax matrix in the poles.

It is easy to see that the Darboux-Nijenhuis variables $\lambda_{1,2}$ (\ref{q-def2}) are related with the poles $Q_{1,2}$ of the Baker-Akhiezer function by the following point transformation
\bq\label{c-var}
\lambda_{1,2}=-\dfrac{\ii}{2}\,Q_{1,2}^2\,.
\eq
In this case the additional knowledge of the Lax matrix allows us to introduce conjugated to $\lambda_k$ momenta
\bq\label{p-def2}
\mu_k = \dfrac{\ii P_k}{Q_k},\qquad \left.P_k=\dfrac{1}{2\ii}\,\ln B(u)\right|_{u=Q_k},
\eq
where $B(u)$ is the diagonal element of the corresponding Lax matrix \cite{ts02}
\ben
B(u)&=&-(x_1-\ii\,x_2)u^3+\Bigl(\bigl(2J_3-\ii\sqrt{c_2}(x_1-\ii x_2)\bigr)(x_1-\ii x_2)-(J_1-\ii J_2)x_3\Bigr)u^2\nn\\
&+&\left(
2x_3J_3(J_1-\ii J_2)+\left(J_1^2+J_2^2+\dfrac{c_4}{x_3^2}\right)(x_1-\ii x_2)-c_2(x_1+\ii x_2)x_3^2
\right)u\nn\\
&+&\ii c_2^{3/2}x_3^4+c_2x_3^3(J_1+\ii J_2)+\ii\sqrt{c_2}x_3^2(J_1-\ii J_2)^2+x_3\left(J_1^2+J_2^2+\dfrac{c_4}{x_3^2}\right)(J_1-\ii J_2)\,.\nn
\en
Moreover, it allows us to prove that variables of separation $(\lambda_{i},\mu_{i})$ lie on the Jacobian of the other hyperelliptic curve of genus two
\bq\label{curv-hyp}
\widetilde{\mathcal C}:\qquad {\Phi}(z,\lambda)=
z^2+\Bigl(4\lambda^2+4\ii H_1\lambda-H_2\Bigr) z +(2a^2\lambda+\ii\,c_4)^2c_2^2=0\,,
\eq
where we put $z=-\ii\sqrt{c_2}\ee^{2\mu\sqrt{2\ii\lambda}}$.

So, we know the answer and, therefore, we could try to guess how to get this information without using our knowledge of the Lax matrix. Namely, the normalized left eigenvectors of $F$ form the St\"ackel matrix
\[F=S^{-1}\,\mbox{diag}(\lambda_1,\lambda_2)\,S,\qquad
S=\left(
 \begin{array}{cc}
  -4\ii q_1 & -4\ii q_2 \\
  1 & 1 \\
 \end{array}
 \right)\,.
\]
In order to get conjugated to $\lambda_{1,2}$ momenta $\mu_{1,2}$ we could use the St\"ackel potentials (\ref{st-pot})
\[
U_{1,2}=-4\ii \lambda_{1,2}H_1+K_2,\quad\mbox{such that}\quad \{U_1,U_2\}=0,\quad \{U_i,\lambda_j\}=0,\qquad i\neq j\,.
\]
As it has been stated above we could study the following recurrence chain of the Poisson brackets
\bq\label{rrel2}
\phi_1=\{\lambda_1,U_1\},\qquad \phi_2=\{\lambda_1,\phi_1\},\ldots,\quad \phi_i=\{\lambda_1,\phi_{i-1}\}.
\eq
This chain is a quasi-periodic chain \[\phi_3=8 \ii \lambda_1 \phi_1.\]
It means that the St\"ackel potential $U_1$ is some trigonometric function on momenta $\mu_1$ and, therefore, we could define this desired momenta in the following way
\[
\mu_1=f(\lambda_1)\ln\Bigl(\sqrt{8\ii \lambda_1}\,\phi_1+\phi_2\Bigr)
\]
up to canonical transformations $\mu_1\to \mu_1+g(\lambda_1)$. Here function $f(\lambda_1)$ is determined from the canonicity of the bracket $\{\lambda_1,\mu_1\}=1$.

Summing up, the central idea of the proposed construction is the observation of quasi-periodicity or of the break of the recurrence chain of the Poisson brackets (\ref{rrel1}) or (\ref{rrel2}).
We have to point out again that all the appropriate tedious calculations may be done on the personal computer in a few seconds. So, it is a real way to get the variables of separation and the separated relations for the given integrable system.

\section{The quadratures}
According to the Liouville theorem the existence of $n$ independent integrals of motion $H_i$
in the involution $\{H_i,H_j\}=0$ guarantees that equations of motion may be solved in quadratures. Usually we fit some additional requirements on these quadratures, as for example, the solutions of equations of motion have to be single valued real functions on the real time variable $t$.
It is necessary for the qualitative analysis of motion, for topological analysis, for perturbation theory etc.

Of course, we prefer to have real solutions represented by an explicit closed-form expression right away. However, in nine times out of ten in order to get such single-valued solutions we have to start with the solution of the Jacobi inversion problem. For example, for the geodesic motion on an ellipsoid and for the Kowalevski top we have to find variables of separation $s_{1,2}(t)$ from the equations
\ben
t+\beta_1 &=& \int_{\infty}^{s_1} \frac{d s}{\sqrt{\mathrm P(s)}}
  +\int_{\infty}^{s_2} \frac{d s}{\sqrt{\mathrm P(s)}},
\nn\\
\beta_2&=& \int_{\infty}^{s_1} \frac{s d s}{\sqrt{\mathrm P(s)}}
  +\int_{\infty}^{s_2} \frac{s d s}{\sqrt{\mathrm P(s)}},\nn
\en
where $\mathrm P$ is a polynomial of degree $5$ or $6$,
and $\beta_{1,2}$ are two constants of integration. Secondly, we have to express the initial real variables via variables of separation $s_{1,2}(t)$.

In the previous section we get the real variables of separation $(q,p)$ and the complex variables of separation $(\lambda,\mu)$ or $(Q,P)$. These variables are related by the canonical transformation
\[
\lambda_{1,2}=\lambda_{1,2}(q_1,q_2,p_1,p_2), \qquad \mu_{1,2}=\mu_{1,2}(q_1,q_2,p_1,p_2)
\]
which may be obtained directly from the definitions (\ref{q-def2},\ref{p-def2}) and the mapping (\ref{inv-map}).
Using the separation relations (\ref{curve-2}) we can rewrite this transformation as a quasi-point canonical transformation \cite{ts96}
\[
\lambda_{1,2}=\lambda_{1,2}(q_1,q_2,H_1,H_2)
\]
which relates the Jacobian of the hyperelliptic curve $\mathcal C$ (\ref{curve-2}) with the Jacobian of the hyperelliptic curve $\widetilde{\mathcal C}$ (\ref{curv-hyp}). However, we can not rewrite it as a rational mapping $\psi:\,(p,q)\to(\lambda,\mu)$, i.e. it does not be well-studied cover of the hyperelliptic curve of genus two even at $c_4=0$, see \cite{fk09} and references within.
We suppose that these Jacobians are non-isogeneous in Richelot sense \cite{rich36} as well. Any further inquiry of this relation goes beyond the scope of this paper.

We have no right to escape complex variables and prefer  the real variables of separation only.
It is how these variables are employed that determines the good or evil. As often as not the equations of motion are linearized on the abelian variety, which is roughly spiking the \textit{complexified} of the corresponding Liouville real torus, see more detailed discussion in \cite{dub}. Moreover, we remind that Lyapunov could improve the Kowalevski result and solve the problem pointed out by Painlev\'{e} by using the \textit{complex} time variable $t$ \cite{lyap}. The Ziglin method \cite{zig}, the theory of algebraically integrable systems \cite{pv} and some other modern theories deal with the \textit{complex} analytical Hamiltonian systems only. Quantum mechanics is formulated over the \textit{complex} field as well.

\subsection{The real variables of separation}
In this case equations of motion reads as
\ben
\dfrac{dq_1}{dt}&=&8q_1(c_2-q_1)p_1\left(1-\dfrac{c_4(q_1-q_2)}
{(4q_1(c_2-q_1)p_1^2-4q_2(c_2-q_2)p_2^2-(q_1-q_2)a^2)}\right),\nn\\
\label{q-eq}\\
\dfrac{dq_2}{dt}&=&8q_2(c_2-q_2)p_2\left(1+\dfrac{c_4(q_1-q_2)}
{(4q_1(c_2-q_1)p_1^2-4q_2(c_2-q_2)p_2^2-(q_1-q_2)a^2)}\right)\nn.
\en
At $c_4=0$ we have an accidental degeneracy of the genus two algebraic curve (\ref{curve-2}) to a product of two elliptic curves (\ref{ell-curve}). Namely, if $c_4=0$ then
\[
\dfrac{dq_1}{8q_1(c_2-q_1)p_1}-\dfrac{dq_2}{8q_2(c_2-q_2)p_2}=2dt
\]
and we have to find functions $q_{1,2}(t,\alpha_{1,2},\beta_{1,2})$ from the two independent equations
\[
\int_{\infty}^{q_1} \frac{d q}{\sqrt{\mathrm P_1(q)}}=\beta_1+t,\qquad\mbox{and}\qquad
\int_{\infty}^{q_2} \frac{d q}{\sqrt{\mathrm P_2(q)}}=\beta_2-t.
\]
Here we fix the values of the integrals of motion $H_{1,2}=\alpha_{1,2}$, and
\[
\mathrm P_{k}(q)=8q_{1,2}(c_2-q_{1,2})\Bigl(a^2(c_2-2q_{1,2})+\alpha_1\mp\sqrt{\alpha_2}\,\Bigr)
\]
are the cubic polynomials defining two elliptic curves (\ref{ell-curve}). According to \cite{chap} in this case we can get solutions of $x_k(t)$ and $J_k(t)$ in an explicit and closed form using the Weierstrass $\wp$ function and its derivative.

At $c_4\neq 0$ we have to solve the standard system of the Abel-Jacobi equations
\[
\int^{q_1}_{\infty}\dfrac{dq}{16q(c_2-q)\,\mathrm P(q)}+\int^{q_2}_{\infty}\dfrac{dq}{16q(c_2-q)\,\mathrm P(q)}=\beta_1+t\,,
\]
and
\ben
& &\int^{q_1}_{\infty}\dfrac{dq}{32q(c_2-q)\Bigl(a^2(c_2-2q)+\alpha_1-8q(c_2-q)\mathrm P^2(q)\Bigr)\mathrm P(q)}\nn\\
&+&\int^{q_2}_{\infty}\dfrac{dq}{32q(c_2-q)\Bigl(a^2(c_2-2q)+\alpha_1-8q(c_2-q)\mathrm P^2(q)\Bigr)\mathrm P(q)}=\beta_2\,.\nn
\en
where $\mathrm P(q)$ means the solution of the equation $\Phi(q,p)=0$ (\ref{curve-2}) with respect to $p$. For example, we can solve this equations numerically by using the Richelot approach \cite{rich36}.

The second part of the Jacobi separation of variables method consists of the construction of the integrable system starting with some known separated variables and some arbitrary separated relations \cite{ts09}. As an example, if we consider new separation relations
\[
\widehat{\Phi}(p,q)=\Phi(p,q)-c_5q^2=0,
\]
where $\Phi(p,q)$ is given by (\ref{curve-2}), then we get the following generalization of the Chaplygin system
\[
\widehat{H}_1=H_1-\dfrac{c_5}{4x_3^4}(J_1^2+J_2^2+c_2x_3^2)\,.
\]
We could not get any physically interesting systems by using these real variables of separation.

\subsection{The complex variables of separation}
In this section we will work with variables $(P,Q)$ instead of $(\mu,\lambda)$, i.e. we will consider a ramified two-sheeted covering given by point transformation (\ref{c-var}).

The poles of the Baker-Akhiezer function lie on the Jacobian of the $2\times 2$ Lax matrix spectral curve \cite{ts02}. This curve is defined by the equation with the real coefficients
\bq
\Phi(y,u)=y^2-(u^4-2H_1u^2+H_2)y+c_2(a^2u^2-c_4)^2=0.
\eq
However, in order to get the separation relations we have to substitute the complex functions $Q_{1,2}$ and $P_{1,2}$ on the initial real variables into this equation
\[
u=Q_{1,2},\qquad y=\ee^{2\ii P_{1,2}}\,.
\]
The Abel-Jacobi equations have the standard form
\[
t+\beta_1 = \int_{\infty}^{Q_1} \Omega_1
  +\int_{\infty}^{Q_2} \Omega_1,
\qquad
\beta_2= \int_{\infty}^{Q_1} \Omega_2
  +\int_{\infty}^{Q_2} \Omega_2,\nn
\]
where
\[
\Omega_1=\dfrac{\partial \Phi(y,u)/\partial H_1}{\partial \Phi(y,u)/\partial y}\,du\qquad \mbox{and}\qquad
\Omega_2=\dfrac{\partial \Phi(y,u)/\partial H_2}{\partial \Phi(y,u)/\partial y}\,du\,.
\]
After the solution of the Abel-Jacobi equations we have to express the real initial variables $x,J$ in terms of this complex solutions $Q_{1,2}(t)$ and $P_{1,2}(t)$.

On the other hand, these complex variables of separation are very useful for the construction of other integrable systems. Namely, substituting $Q_{1,2}$ and $P_{1,2}$
into other separation relations we can get the Hamilton function for the Ko\-wa\-lev\-ski-Goryachev-Chaplygin gyrostat \cite{kuzts89,ts02}
\bq\label{g-Ham}
\widehat{H}_1=J_1^2+J_2^2+2J_3^2+\rho J_3+c_1 x_1+c_2(x_1^2-x_2^2)+c_3x_1x_2+\dfrac{c_4}{x_3^2}
\eq
after some additional canonical transformations. Moreover, using these complex variables we can easily get quantum counterpart of the Chaplygin system and its generalizations \cite{kuzts89}.

\section{Conclusion}

Starting with integrals of motion for the generalized Chaplygin system we have found the two polynomial in momenta Poisson bivectors (\ref{P2}) and (\ref{P3}), which are compatible with the canonical Poisson bivector on cotangent bundle $T^*\mathcal S^2$ of the two-dimensional sphere.

An application of the corresponding control matrices allows us to get two families  variables of separation and of separated relations using methods of bi-hamiltonian geometry only. The solutions of equations of motion in these variables are briefly discussed.

The proposed approach may be useful for investigations of other integrable systems on the sphere with integrals of motion higher order in momenta \cite{yeh08}, for instance,  the search another real variables for the Ko\-wa\-lev\-ski-Goryachev-Chaplygin gyrostat (\ref{g-Ham}).


\begin{thebibliography}{10}




\bibitem{chap}
S.A. Chaplygin,
 \newblock{\em A new partial solution of the problem of motion of a rigid body in a liquid}, Trudy otdel. Fiz . Nauk Obsh. Liub. Est. v.11, p.7–10, 1903.

\bibitem{dub} B. A. Dubrovin, 	
\newblock{\em Riemann Surfaces and Nonlinear Equations},
AMS, 2002.

\bibitem{fp02}
G. Falqui, M. Pedroni,
\newblock{\em Separation of variables for bi-Hamiltonian systems},
\newblock{Math. Phys. Anal. Geom.}, v.6, p.139-179, 2003.

\bibitem{fk09}
G. Frey, E. Kani,
\newblock{\em Curves of genus 2 with elliptic differentials and
associated Hurwitz spaces},
In: Arithmetic, Geometry, Cryptography and Coding Theory (Lachaud et al, eds.) Contemp. Math. v.487, pp.33-81, 2009.


\bibitem{kuzts89}
V.B. Kuznetsov, A.V. Tsiganov, \newblock{\em A special case of
Neumann's system and the Kowalewski-Chaplygin-Goryachev top},
 {J. Phys. A.}, v.22, p.L73-79, 1989.

\bibitem{ts96}
 S. Rauch-Wojciechowski, A.V. Tsiganov, \newblock{\em
Quasi-point separation of variables for H\'{e}non-Heiles system and system with quartic potential},
Journal of Physics A, v.29, p.7769--7778, 1996.



\bibitem{lyap}
A.M. Lyapunov,\newblock{\em On a Certain Property of the Differential
Equations of the Problem of Motion of a Heavy
Rigid Body Having a Fixed Point}, Soobshch. Kharkov
Math. Obshch., Ser. 2, v.4, pp. 123-140, 1894. Collected
Works, V. 5, Izdat. Akad. SSSR, Moscow, 1954, (in
Russian).

\bibitem{mag97}
F. Magri, \newblock{\em Eight lectures on Integrable Systems.}
In: Integrability of Nonlinear Systems (Y. Kosmann-Schwarzbach et
al. eds.), Lecture Notes in Physics {\bf 495}, Springer Verlag,
Berlin-Heidelberg, 1997, pp.\ 256--296.

\bibitem{rich36}
 F. Richelot,\newblock{\em Essai sur unem\'{e}thode g\'{e}n\'{e}rale pour d\'{e}terminer la valeur des int\'{e}grales ultra-elliptiques, fond\'{e}e
sur des transformations remarquables de ces transcendantes}, C. R. Acad. Sci., Paris, v.2, pp.622–627, 1836.

\bibitem{ts02}
A.V. Tsiganov,	\newblock{\em On the Kowalevski-Goryachev-Chaplygin gyrostat},
J. Phys. A, Math. Gen. v.35, No.26, L309-L318, 2002.

\bibitem{ts07a}
A.V. Tsiganov,	\newblock{\em On the two different bi-Hamiltonian structures for the Toda lattice}, Journal of Physics A: Math. Theor. v.40, pp. 6395-6406, 2007.


\bibitem{ts07c}
A.V. Tsiganov, \newblock{\em Separation of variables for a pair of integrable systems on $so^*(4)$}, Doklady Math., v.76(3), p.839-842, 2007.

\bibitem{ts08}
A.V. Tsiganov,
\newblock{\em On bi-hamiltonian structure of some integrable systems on $so^*(4)$},
J. Nonlinear Math. Phys., v.15(2), p.171-185, 2008.


\bibitem{ts08b}
A.V. Tsiganov,
\newblock{\em On bi-hamiltonian geometry of the Lagrange top},
J. Phys. A: Math. Theor., v.41, 315212 (12pp), 2008.

\bibitem{ts08c}
A.V. Tsiganov,
\newblock{\em The Poisson bracket compatible with the classical reflection equation algebra},
Regular and Chaotic Dynamics, v.13(3), 191-203, 2008.

\bibitem{pv}
P. Vanhaecke,\newblock{\em Integrable Systems in the Realm of
Algebraic Geometry}, Lecture Notes in Math. 1638.
Springer-Verlag, Berlin, 1996.


\bibitem{ts09}
A.V. Vershilov, A.V. Tsiganov,
\newblock{\em
On bi-Hamiltonian geometry of some integrable systems on the sphere with cubic integral of motion}, J. Phys. A: Math. Theor. v.42, 105203 (12pp), 2009.

\bibitem{yeh08}
H.M. Yehia, A.A. Elmandouh,\newblock{\em New Integrable Systems with a Quartic Integral and New Generalizations of Kovalevskaya's and Goriatchev's Cases},
Regular and Chaotic Dynamics, v.13(1), pp.56 - 69 , 2008.

\bibitem{zig}
S. L. Ziglin,\newblock{\em Branching of Solutions and Non-Existence
of First Integrals in Hamiltonian Mechanics I,II}. Funct.
Anal. Appl., v.16, pp. 181-189, 1982, v.17, pp. 6-17, 1983.





\end{thebibliography}
\end{document}